\documentclass[12pt]{article}
\usepackage{a4wide}
\usepackage{amssymb}
\usepackage{amsmath}
\usepackage{mathrsfs}
\usepackage{bm}
\usepackage{graphicx} 
\begin{document}
\newcommand{\be} {\begin{equation}}
\newcommand{\ee} {\end{equation}}
\newcommand{\ba} {\begin{eqnarray}}
\newcommand{\ea} {\end{eqnarray}}
\newcommand{\e} \epsilon
\newcommand{\la} \lambda
\newcommand{\La} \Lambda

\author{B. Boisseau\thanks{E-mail :
bruno.boisseau@lmpt.univ-tours.fr}\\
\small Laboratoire de Math\'ematiques et Physique Th\'eorique\\
\small CNRS/UMR 6083, Universit\'e Fran\c{c}ois Rabelais\\
\small
\small F\'ed\'eration Denis Poisson\\
\small Parc de Grandmont 37200 TOURS, France}

\title{\bf Exact
 cosmological
 solution of a Scalar-Tensor Gravity theory compatible  with the $\Lambda CDM$ model}
\date{}

\maketitle

\begin{abstract}
We consider the massive scalar-tensor theory in the Jordan frame $F(\Phi) =K^{2}\Phi^2$ and $U(\Phi) =(1/2)m^{2}\Phi^2$, where $F(\Phi)$ corresponds
to a constant Brans-Dicke parameter $\omega_{BD}=1/4K^2$. The constraint of the Solar System experiments is $K^2<(1/400)^2$. For dustlike matter in a spatially flat homogeneous isotropic universe, we reduce the equations of motion to a system of two differential equations of first order   which can be exactly solved.  We obtain simple and explicit expressions for $\frac{\Phi(z)}{\Phi(0)}$ and $\frac{H(z)}{H_{0}}$ that depend only on two parameters, $K^2$ and $\Omega_{m,0}$.  For  $K\leq1/400$ the expansion rate $H(z)$  can be practically superposed on the $\Lambda$CDM solution  $H_{\Lambda}(z)$, 
 up to high redshift $z$, but the equation of state  $w_{DE}(z)$ of the dark energy is not constant: it presents a very slight crossing of the phantom divide line $w=-1$ in the neighborhood of $z=0$ and becomes very slightly positive at high redshifts.


\end{abstract}

\section{Introduction}

Supernova observations \cite{perlmutter}, \cite{riess98},  \cite{riess-gold}, \cite{astier06} provide  strong evidence for a late-time accelerated expansion of our  Universe. These data, combined with the observations of the microwave background and gravitational clustering,   strongly suggest that our universe is spatially flat and composed of about two-thirds of an unknown form of negative pressure unclustered matter called Dark Energy (DE) \cite{sahni-notes-2004}, \cite{spergel et al03}, \cite{tegmark et al04}.
 The Friedmann-Robertson-Walker equations with a cosmological constant $\Lambda$ as source of dark energy are the simplest model (called $\Lambda CDM$), which fits remarkably well with the current cosmological observations \cite{perivol-10}. However it is difficult to explain the magnitude of $\Lambda$, which is unnaturally small (e.g. \cite{silvestri-trodden}). In the $\Lambda CDM$ model the energy density of the  dark energy  $\rho_{DE}=\frac{\Lambda}{8\pi G}$ and the pression $P_{DE}=-\rho_{DE}$ are constants, so that the value of $w_{DE}=\frac{P_{DE}}{\rho_{DE}}$ is exactly $-1$, that is $w_{DE}$ is exactly on the phantom divide line. The uncertainty on the observations tells us only that  $w_{DE}(z)$ is close to $-1$ and can eventually cross the phantom divide line at a late time. So there is  intense activity on models likely to cross the divide line; among them, the class of the scalar-tensor theories, which is the subject of this paper.
 
Since there is no obvious reason to choose a particular model, defined by the functions $F(\Phi)$ and $U(\Phi)$, in many works on the subject the phenomenological aspects are  emphasized, i.e. on  reconstructing these functions from observations. The knowledge of the luminosity distance $ D_{L}(z)$ and the linear density perturbation $\delta_{m}(z)$ is sufficient to reconstruct the full theory \cite{beps}. It is  also possible  to consider a restricted class, by giving an additional constraint {\em a priori}, for example by giving $F(\Phi)$ and reconstructing  $U(\Phi)$ from the  knowledge of the luminosity distance ($ D_{L}(z)$  \cite{ep2000} or by giving a parametrization of  $w_{DE}(z)$ or $ H(z) $ and 
reconstructing   $U(\Phi)$ and  $F(\Phi)$, \cite{sahni-star2006}, \cite{perivol-jcap-05}.  
  
 In this paper we consider  the  model  
 $F(\Phi) =K^2\Phi^2$ and $U(\Phi) =(1/2)m^2\Phi^2$. 
 Its particular appeal lies in that  the Brans-Dicke parameter 
$\omega_{BD}\equiv F(\Phi)/(\frac{dF}{d\Phi})^2$  has a constant value $1/4K^2$
and $U(\Phi)$ is the standard mass potential. 
 From the Solar System experiments, $\omega_{BD}\geq 40000$, hence $K^{2}\leq(1/400)^2$ is a small parameter. For dustlike matter in a spatially flat, homogeneous and isotropic universe, we obtain an exact solution for the expansion rate $ H(z) $,  compatible with its $\Lambda CDM$ counterpart up to a large redshift.

Let us note that this particular  scalar-tensor model has been studied several times in different contexts. In particular, under the same conditions, that is for dustlike matter in a spatially flat homogeneous isotropic universe, K. Uehara and C. W. Kim \cite{uehara-kim} have solved this system (in the Brans-Dicke form) and given an exact solution for the scale factor $ a(t) $. But this function of time is not directly adapted for comparison with experiments which require functions of the redshift $ z $, and it seems very difficult  to extract an exact $ H(z) $ from their solution $ a(t) $. 
 
 In section $2$, we introduce the equations of motion for a scalar-tensor theory in a Jordan frame for dustlike matter in a spatially flat,  homogeneous and isotropic universe and carry out some preliminary calculations that set the stage for our approach.
In section $ 3 $ we derive, from the equations of motion, a system of two differential equations of first order for  $ u(z)=(H(z)/H_{0})^2 $ and $ S(z)=\frac{d\Phi/dz}{\Phi(z)} $ where the independent variable is the redshift $ z $ . In section $ 4 $, we obtain an exact solution of this differential system. The quantities $h(z)=\frac{H(z)}{H_{0}}$  and  $\frac{\Phi(z)}{\Phi(0)}$ are explicitly expressed as functions of the two parameters $K^{2}$ and $\Omega_{m,0}$ and $ h(z) $ is  compared with its $\Lambda CDM$ counterpart. In section $ 5 $, we look at the equation of state  $w_{DE}(z)$ of the dark energy, which is not constant and we show that it displays a slight crossing of the phantom divide line $w=-1$ in the neigbourhood of $z=0$ and tends to $ \frac{4}{3}K^2+O(K^4) $ at infinity. We show that the dark energy is negligible at high redshift.
 In section $ 6 $, by a local dynamical analysis, we show that the orbit of this solution approaches a stable proper node. In section $ 7 $, we conclude and explain why it would be interesting to fit the solution $ H(z) $ to the data of the supernovae observations.

\section{ Scalar-Tensor Gravity and  $\Lambda$CDM}
Let us consider a universe where gravity is described by a scalar-tensor theory. The Lagrangian density in Jordan frame is given by 

\be
\label{2.1}
L= \frac{1}{2}\left( F(\Phi)R-g^{\mu\nu}\partial{\mu}\Phi\partial{\nu}\Phi-2U(\Phi)\right)+L_{m}(g_{\mu\nu}) ,
\ee
where $L_{m}(g_{\mu\nu})$ describes the dustlike matter.   

For a spatially flat homogeneous isotropic  universe whose metric is given by

\be
\label{2.2}
ds^2=-dt^2+a^2(t)d\vec{x}^2 ,
\ee
the equations of the motion are
\be
\label{2.3}
3FH^2=\rho_{m}+\frac{1}{2}\dot{\Phi}^2-3H\dot{F}+U,
\ee

\be
\label{2.4}
-2F\dot{H}=\rho_{m}+\dot{\Phi}^2+\ddot{F}-H\dot{F},
\ee

\be
\label{2.5}
\ddot{\Phi}+3H\dot{\Phi}=3\frac{dF}{d\Phi}(\dot{H}+2H^2)-\frac{dU}{d\Phi},
\ee
where $H=\frac{\dot{a}}{a}$ describes the evolution of the expanding universe and  $\rho_{m}$ is the energy density of the dust matter. It is useful to have also the equation of conservation of matter
\be
\label{2.6}
\dot{\rho}_{m}+3H\rho_{m}=0,
\ee
which is not independent of the other equations.

It is convenient to introduce the dimensionless quantities
\be
\label{2.9}
h=\frac{H}{H_{0}} \quad , \quad t'=H_{0}t
\ee
where $H_{0}=H(t_0)$ is the Hubble constant.
The quantities evaluated today at $t_{0}$ will be denoted with a zero index.

We successively eliminate $\rho_{m}$ from the equations (\ref{2.3}) and (\ref{2.4}), and  $\ddot{\Phi}$ from the resulting equation and the equation (\ref{2.5}), then writing these equations, using the dimensionless quantities (\ref{2.9}), we obtain a  system of two differential equations for $h(t')$ and $\Phi(t')$   

\be\label{2.11}
h^2(1+2\frac{(F_{,\Phi})^{2}}{F})+\frac{2}{3}h'(1+\frac{3}{2}\frac{(F_{,\Phi})^{2}}{F})+\frac{1}{3F}(\frac{1}{2}+F,_{\Phi^2})\Phi'^{2}-\frac{F_{,\Phi}h\Phi'}{3F}-(\frac{F_{,\Phi}U_{,\Phi}}{3FH^{2}_{0}}+\frac{U}{3FH^{2}_{0}})=0,
\ee
\be
\label{2.12}
\Phi''+3h\Phi'+\frac{U_{,\Phi}}{H_{0}^2}-3F_{,\Phi}(h'+2h^2)=0,
\ee
where $ \Phi'=\frac{d\Phi}{dt'} $ etc...
Another equation will be useful, from (\ref{2.3}) at $t'_{0}$:
\be
\label{2.13}
1=\Omega_{m,0}+\frac{(\Phi'_{0})^2}{6F_{0}}+\frac{U_{0}}{3F_{0}(H_{0})^2}-\frac{1}{F_{0}}(\frac{\partial F}{\partial\Phi})_0\Phi'_{0},
\ee
where
\be
\label{2.14}
\Omega_{m,0}=\frac{\rho_{m,0}}{3F_{0}H_{0}^2}.
\ee

As indicated in the introduction, we choose now
\be
\label{2.15}
F=K^2\Phi^2 \quad ,\quad U=\frac{1}{2}m^2\Phi^2.
\ee 
The equations (\ref{2.11}), (\ref{2.12}), (\ref{2.13}) become
\be
\label{2.16}
(1+8K^2)h^2+\frac{2}{3}(1+6K^2)h'+(\frac{1}{6K^2}+\frac{2}{3})\frac{\Phi'^2}{\Phi^2}-\frac{2}{3}h\frac{\Phi'}{\Phi}-\frac{1}{6}(4m_0^2+\frac{m_0^2}{K^2})=0,
\ee

\be
\label{2.17}
\Phi''+3h\Phi'+m_0^2\Phi-6K^2\Phi(h'+2h^2)=0,
\ee
\be
\label{2.18}
1=\Omega_{m,0}+\frac{(\Phi'_0)^2}{6K^2(\Phi_0)^2}+\frac{m_0^2}{6K^2}-\frac{2\Phi'_0}{\Phi_0},
\ee
with
\be
\label{2.19}
m_0=\frac{m}{H_0}.
\ee

It will be interesting to compare these equations to those of the corresponding  $\Lambda$CDM model that we recall here:
\be
\label{2.20}
H^2=\frac{8\pi G \rho_{m}}{3}+\frac{\Lambda}{3},
\ee
\be
\label{2.21}
\frac{\ddot{a}}{a}=-\frac{8\pi G \rho_{m}}{6}+\frac{\Lambda}{3}.
\ee
From (\ref{2.20}) and (\ref{2.21}), we obtain
\be
\label{2.22}
h^2+\frac{2}{3}h'-\Omega_{\Lambda,0}=0,
\ee
\be
\label{2.22'}
1=\Omega_{m,0}+\Omega_{\Lambda,0},
\ee
with
\be
\label{2.23}
\Omega_{m,0}=\frac{8\pi G \rho_{m,0}}{3H_{0}^2}\quad,\quad\Omega_{\Lambda,0}=\frac{\Lambda}{3H_0^2}.
\ee

Let us note that the convention chosen in the  Lagrangian (\ref{2.1}) implies  
$8\pi G=1/F_0$ to very high accuracy.

\section{Reduction to a system of two differential equations of the first order}
The redshift $z$ is given by 
\be
\label{2.8}
1+z=\frac{a_{0}}{a}.
\ee
From  (\ref{2.8}) and (\ref{2.9}) we have

\be
\label{3.1}
dt'=-\frac{dz}{(1+z)h(z)},
\ee
and the differential equations  (\ref{2.16}) and (\ref{2.17}) are rewritten in terms of the variable $ z $:
\be
\label{3.2}
\begin{array}{l}
\displaystyle
h^2(1+8K^2)-\frac{2}{3}(1+6K^2)(1+z)h\frac{dh}{dz}+(\frac{1}{6K^2}+\frac{2}{3})(1+z)^2h^2\frac{(d\Phi/dz)^2}{\Phi^2}\\
\displaystyle
+\frac{2}{3}(1+z)h^2\frac{(d\Phi/dz)}{\Phi}-\frac{1}{6}(4m_0^2+\frac{m_0^2}{K^2})=0,
\end{array}
\ee
\be
\label{3.3}
(1+z)^2\left( h\frac{dh}{dz}\frac{d\Phi}{dz}+h^2\frac{d^2\Phi}{dz^2}\right) -2(1+z)h^2\frac{d\Phi}{dz}+m_0^2\Phi+6K^2\Phi\left( (1+z)h\frac{dh}{dz}-2h^2\right) =0.
\ee

We now express the system  (\ref{3.2}), (\ref{3.3}) as a system of the first order in the new functions 
\be
\label{3.3'}
 u(z)=h^2(z)  \quad , \quad S(z)=\frac{d\Phi/dz}{\Phi} :
\ee
\be
\label{3.5}
\begin{array}{l}
\displaystyle
-\frac{1}{3}(1+6K^2)(1+z)\frac{du}{dz}+\left( 1+8K^2+\frac{2}{3}(1+z)S(z)+(\frac{2}{3}+\frac{1}{6K^2})(1+z)^2S^2(z)\right) u(z)\\
\displaystyle
-\frac{m_0^2}{6}(4+\frac{1}{K^2}) =0,
\end{array}
\ee
\be
\label{3.6}
\begin{array}{l}
\displaystyle
\left( \frac{1}{2}(1+z)^2 S(z)+3K^2(1+z)\right)\frac{du}{dz}+\\
\displaystyle
\left(-12K^2-2(1+z)S(z)+(1+z)^2 \left( \frac{dS}{dz}+S^2(z)\right) \right)u(z)                                                                           
+m_0^2 =0.
\end{array}
\ee

\section{Exact solution and comparison with  $\Lambda$CDM}
We remark that the two equations (\ref{3.5}), (\ref{3.6}) of the differential system are written in the same form as first order linear equations in $ u(z) $: 
\be
\label{4.1}
A_i\frac{du}{dz}+B_iu(z)+C_i=0 \quad , \quad i=1,2,
\ee
where the coefficients $ A_i,B_i,C_i $ depend in general on $ z,S(z),\frac{dS}{dz} $, and 
on the parameters $ K^2,m_0^2 $. 

In order to have a solution $ u(z) $ of this system it is sufficient to find $ S(z) $ which ensures the proportionality between the coefficients: 
\be
\label{4.2}
A_2=\mu A_1 \quad , \quad B_2=\mu B_1 \quad , \quad C_2=\mu C_1,
\ee
that is:
\be
\label{4.3}
\left( \frac{1}{2}(1+z)^2 S(z)+3K^2(1+z)\right)=\mu (-\frac{1}{3}(1+6K^2)(1+z))
\ee

\be
\label{4.4}
\begin{array}{l}
\displaystyle
\left(-12K^2-2(1+z)S(z)+(1+z)^2 \left( \frac{dS}{dz}+S^2(z)\right) \right)=\\
\displaystyle
\mu\left( 1+8K^2+\frac{2}{3}(1+z)S(z)+(\frac{2}{3}+\frac{1}{6K^2})(1+z)^2S^2(z)\right)
\end{array}
\ee

\be
\label{4.5}
m_0^2=\mu(\frac{-m_0^2}{6})(4+\frac{1}{K^2})
\ee
Fortunately, the system of equations (\ref{4.3}), (\ref{4.4}) and (\ref{4.5}) has a solution $ (\mu,S(z)) $.

The solution of (\ref{4.5}) gives
\be
\label{4.6}
\mu=-(\frac{6K^2}{1+4K^2})
\ee
that we insert in (\ref{4.3}), which is an algebraic equation in $ S(z) $. We obtain

\be
\label{4.7}
S(z)=-\frac{2K^2}{(1+4K^2)(1+z)}.
\ee
Inserting (\ref{4.6}) and (\ref{4.7}) in the differential equation  (\ref{4.4}), we find that it is identically verified.
 
We remark that the solution (\ref{4.7}) of $ S(z) $ is independent of $ m_0 $ and that there is not any unknown constant of integration since it is obtained by solving an algebraic equation.
We have
\be
\label{4.8}
S(0)=-\frac{2K^2}{(1+4K^2)}.
\ee

Now, we can obtain the solution $ u(z) $ by replacing $ S(z) $ given by (\ref{4.7}) in one of the two equations of the system  (\ref{3.5}), (\ref{3.6}). We obtain a linear first order equation in $ u(z) $:
\be
\label{4.9}
(1+4K^2)(1+6K^2)(1+z)\frac{du}{dz} -(3+34K^2+96K^4)u(z) +\frac{m_0^2}{2K^2}(1+4K^2)^2=0 .
\ee  
It will be better for comparison with  $\Lambda CDM$ to introduce the parameter 
$ \Omega_{m,0} $ instead of $ m_0^2 $.
We eliminate $ m_0^2 $ from (\ref{2.18})
\be
\label{4.10}
m_0^2= 6K^2\left[ 1-\Omega_{m,0}-2S(0)\right] -S^2(0)
\ee
where we have used
\be
\label{4.11}
\frac{\Phi'_0}{\Phi_0}=-\frac{(d\Phi/dz)_0}{\Phi_0}=-S(0),
\ee
and with (\ref{4.8}) we write (\ref{4.9}) in the form
\be
\label{4.12}
(1+4K^2)(1+6K^2)(1+z)\frac{du}{dz} -(3+34K^2+96K^4)u(z) +3+34K^2+96K^4-3\Omega_{m,0}(1+4K^2)^2=0 ,
\ee 
whose solution is  determined by the initial condition $ u(0)=1 $ :

\be
\label{4.13}
u(z)= 1-\overline{\Omega}_0 + \overline{\Omega}_0(1+z)^{3+\frac{4K^2}{1+4K^2}},
\ee
where
\be
\label{4.14}
\overline{\Omega}_0 = \Omega_{m,0} \frac{3+24K^2+48K^4}{3+34K^2+96K^4}.
\ee
In the $\Lambda$CDM model, equation (\ref{2.22}) becomes, with the use of (\ref{2.22'}),

\be
\label{4.15}
-\frac{1}{3}(1+z)\frac{du}{dz}+u(z)-(1-\Omega_{m,0})=0,
\ee
whose well-known solution is
\be
\label{4.16}
u_{\Lambda}(z)=1-\Omega_{m,0}+\Omega_{m,0}(1+z)^3.
\ee
Now we know, as indicated in the introduction, that we must have $ K^2\leq 1/(400)^2 $ from solar experiments. So, it is obvious from  (\ref{4.13}), (\ref{4.14}) and  (\ref{4.16}) that solution $ u(z) $ is very close to 
$ u_{\Lambda}(z) $.
 
\begin{figure}[p]
\includegraphics[scale=1]{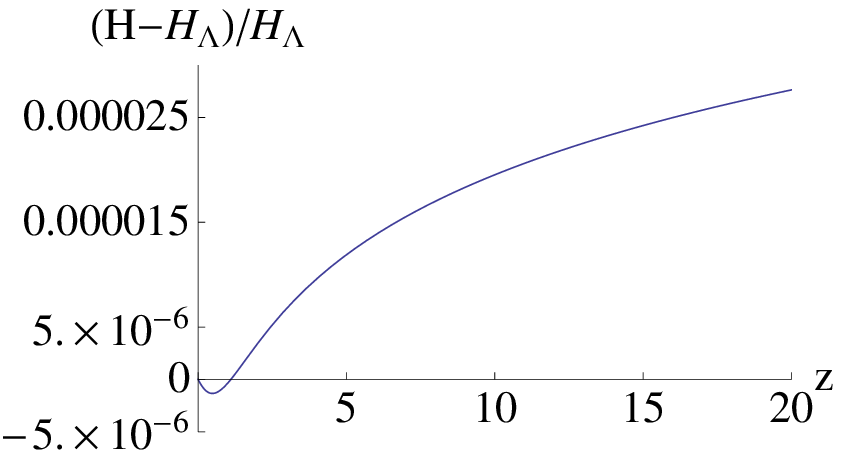}
\caption[]{Proximity of $ H(z)$ and $ H_\Lambda(z) $. $ K=1/400 $, $ \Omega_{m,0}=0.3 $.}
\label{fig1}
\end{figure}

From the exact solution  (\ref{4.13}), (\ref{4.7}) and  (\ref{3.3'}) we have immediately

\be
\label{4.19}
\frac{H(z)}{H_0}=h(z)=\sqrt{ 1-\overline{\Omega}_0 + \overline{\Omega}_0(1+z)^{3+\frac{4K^2}{1+4K^2}}}
\ee
and
\be
\label{4.20}
\frac{\Phi(z)}{\Phi_0}=(1+z)^{-\frac{2K^2}{1+4K^2}}.
\ee
We illustrate in Figure \ref{fig1}, for $ K^2=1/(400)^2 $ 
 and $ \Omega_{m,0}=0.3 $, the proximity of the expansion rate $ H(z) $ with $ H_{\Lambda}(z)$ the expansion rate of the  $\Lambda$CDM model by  plotting  the relative difference $ (H(z)- H_{\Lambda}(z))/ H_{\Lambda}(z)$ which is independent of $ H_0 $.

Let us observe from  (\ref{4.10}) and (\ref{4.8}) that we have an algebraic relation between $  m_0^2 $, $ K^2 $ and $ \Omega_{m,0} $ which can be written
$$ m_0^2= 6K^2(1-\Omega_{m,0})+O(K^4). $$
 This relation shows that  the mass $ m=m_0 H_0 $ is very small, which imposes a strong constraint on the model.

 It is interesting to give an approximation of the exact solution $ u(z) $ to first order in $ K^2 $:
\be
\label{4.17}
u_{approx}(z)= 1-\Omega_{m,0}(1-\frac{10}{3}K^2)+\Omega_{m,0}(1-\frac{10}{3}K^2)(1+z)^3+
4K^2\Omega_{m,0}(1+z)^3\ln (1+z).
\ee 
We see that we can neglect $ K^2=1/(400)^2 $ in the two first terms of (\ref{4.17}) but the last term can dominate for very large $ z $. However this term is still $ 1/100 $ of the last but one, that is $ 4K^2\ln (1+z)= \frac{1}{100} $, for $ z \simeq 5.22 \times 10^{173} $ which is a very large redshift.

For $ z=3000 $, which marks approximately the transition 
between the nonrelativistic matter and the relativistic matter (not taken into account in this model), we have $ 4K^2\ln (1+3000)\simeq 2\times 10^{-4} $; the last term is negligible and we can write 
\be
\label{4.18}
u_{\Lambda}(z)\simeq u_{approx}(z)\simeq u(z).
\ee

In the next section, we shall return to this approximation, in relation with the equation of state parameter of the dark energy 
$ w_{DE}(z) $ .

\section{Equation of state parameter of the dark energy }

\begin{figure}[p]
\includegraphics[scale=1]{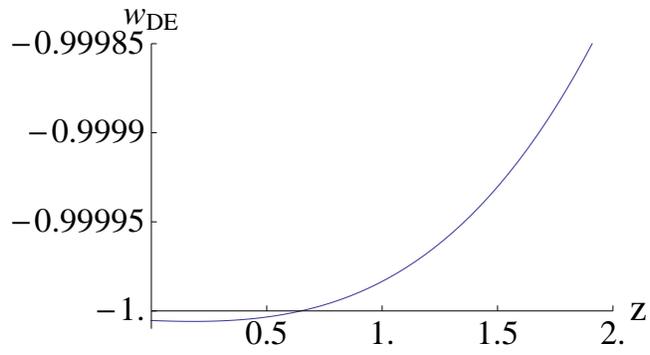}
\caption[]{$ w_{DE}(z)$. $ K=1/400 $, $ \Omega_{m,0}=0.3 $. Slight crossing  of the phantom divide.}
\label{fig2}
\end{figure}

\begin{figure}[p]
\includegraphics[scale=1]{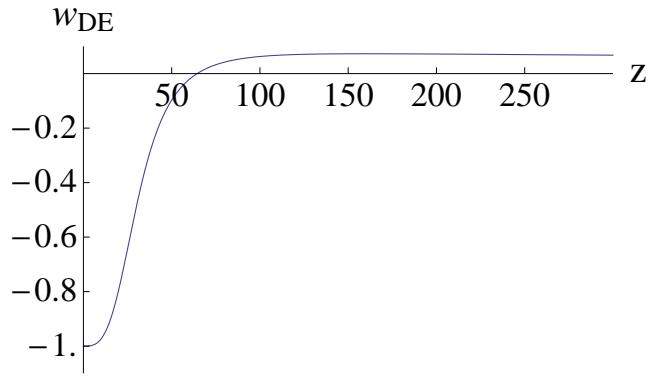}
\caption[]{$ w_{DE}(z)$. $ K=1/400 $, $ \Omega_{m,0}=0.3 $. }
\label{fig3}
\end{figure}

\begin{figure}[p]
\includegraphics[scale=1]{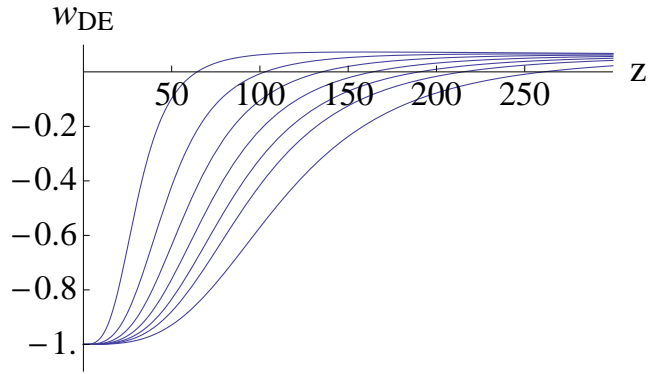}
\caption[]{$ w_{DE}(z)$. $ K=1/400 $ to $ K=1/32000 $, $ \Omega_{m,0}=0.3 $. When $ K $ decreases, $ w_{DE}(z) $ takes more time close to the value $ -1 $. }
\label{fig4}
\end{figure}

We can define the DE density $ \rho_{DE} $ and pression $ P_{DE} $ by writing the equations (\ref{2.3}) (\ref{2.4}) in the form \cite{GPRS2006}
\be
\label{5.1}
3F_0H^2=\rho_{m}+\rho_{DE},
\ee
\be
\label{5.2}
-2F_0\dot{H}=\rho_{m}+\rho_{DE}+P_{DE}.
\ee
With these definitions the usual conservation equation applies:
\be
\label{5.3}
\dot{\rho}_{DE}= -3H(\rho_{DE}+P_{DE}).
\ee
And if we define $ w_{DE}=\frac{P_{DE}}{\rho_{DE}} $, we obtain
\be
\label{5.4}
w_{DE}=\frac{\frac{1}{3}(1+z)\frac{du(z)}{dz}-u(z)}{u(z)-\Omega_{m,0}(1+z)^3}.
\ee
Details are given in  \cite{GPRS2006}.
From  (\ref{5.4}) and  (\ref{4.13}), we write
\be
\label{5.5}
w_{DE}(z)=\frac{-1+\overline{\Omega}_0+\frac{4}{3}\frac{K^2}{1+4K^2}\overline{\Omega}_0(1+z)^{3+\frac{4K^2}{1+4K^2}}}{1-\overline{\Omega}_0+\overline{\Omega}_0(1+z)^{3+\frac{4K^2}{1+4K^2}}-\Omega_{m,0}(1+z)^3},
\ee
hence
\be
\label{5.6}
w_{DE}(\infty)=\frac{4}{3}\frac{K^2}{1+4K^2} \simeq 8\times 10^{-6}.
\ee
We can also see that we have a slight crossing of the phantom line in the neighborhood of 
$ z=0 $. The condition $ w_{DE}(0)<-1 $ is written

\be
\label{5.7}
\frac{1}{3}(\frac{du}{dz})_0 -\Omega_{m,0}<0,
\ee
 which is easily verified  using $ u_{approx}(z) $ given by  (\ref{4.17}):
\be
\label{5.8}
-2K^2\Omega_{m,0}<0.
\ee
The plotted Figures (\ref{fig2}, \ref{fig3}) show $  w_{DE}(z) $ at different scales. Figure (\ref{fig4}) displays the evolution of $  w_{DE}(z)$, when $ K $ decreases from $ K=1/400 $, $  w_{DE}(z) $ takes more time close to the value $ -1 $.

It is now customary to write $ u(z) $  \cite{GPRS2006} as a sum of  terms of nonrelativistic energy and dark energy:
\be
\label{5.81}
u(z)=\Omega_{m,0}(1+z)^3+(1-\Omega_{m,0})\epsilon (z),
\ee
where $ \epsilon (z) $ is expressed as a function of $ w_{DE} (z) $ by
\be
\label{5.82}
\epsilon (z)=\exp \left[ 3 \int_{0}^{z} dz'\frac{1+w_{DE}(z')}{1+z'}\right].
 \ee
In order to evaluate $ \epsilon (z) $, it is simpler to compare  (\ref{5.81}) with  (\ref{4.17}), which is written:
\be
\label{5.83}
u_{approx}(z)\simeq 1-\Omega_{m,0}+\Omega_{m,0}(1+z)^3+
4K^2\Omega_{m,0}(1+z)^3\ln (1+z).
\ee 
(It is legitimate to neglect $ K^2=(1/400)^2 $ in the two first terms.)

Hence, a good approximation of  $ \epsilon (z) $ is given by
\be
\label{5.84}
\epsilon (z)\simeq 1+4K^2\frac{\Omega_{m,0}}{ 1-\Omega_{m,0}}(1+z)^3\ln (1+z).
\ee 

It is immediate to incorporate the result of the end of the last section: The part of the dark energy $ (1-\Omega_{m,0})\epsilon  $ becomes preponderant for very large $ z $ but it is yet $ 1/100 $ of the term of nonrelativistic matter $ \Omega_{m,0}(1+z)^3 $ for the huge redshift $  z \simeq 5.22 \times 10^{173} $ .
At $ z=3000 $ it is perfectly negligible. 
So this term should be significant only at the very beginning of the Universe. But we have not included in our model the relativistic matter $ \sim (1+z)^4 $ which becomes significant at about $ z=3000 $ and can overtake the dark energy $ \sim (1+z)^3\ln (1+z) $ at large $ z $. This  last remark suggests that the dark energy of this model is   never significant at large redshifts.

\begin{figure}[p]
\includegraphics[scale=1]{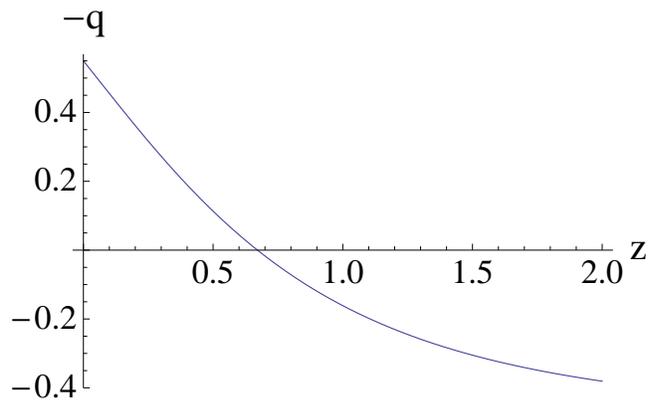}
\caption[]{ acceleration $-q(z) $. $ K=1/400 $, $ \Omega_{m,0}=0.3 $. The 
acceleration begins at $ z\simeq 0.65 $.}
\label{fig5}
\end{figure}

The deceleration parameter $ q=-\frac{\ddot{a}/a}{H^2} $ is given by
\be
\label{5.9}
q(z)=-1+\frac{1}{2}(1+z)\frac{\frac{du}{dz}}{u(z)}.
\ee
The plot of $ -q(z) $ (Figure \ref{fig5}) shows that the acceleration of the Universe in this model ($ \Omega_{m,0}=.3,K=1/400 $) begins at $ z\simeq 0.65 $. It can be verified that $ q(z) $ is virtually independent of $ K $ where $ K $ is compatible with the Solar System experiments.

\section{Dynamical study around the exact solution }

The system of differential equations (\ref{3.5}), (\ref{3.6})  is transformed to an autonomous dynamical system by setting 

\be
\label{6.1}
v=S(z)(1+z),
\ee
and the change of variable
\be
\label{6.2}
p=-\ln (1+z).
\ee
($ p $ is growing as a function of time.)

We obtain:

\be
\label{6.3}
\frac{du}{dp}=\frac{(1+4K^2)m_0^2-u(p)\left( 6K^2+48K^4+4K^2v(p)+(1+4K^2)v^2(p)\right) }{2K^2(1+6K^2)},
\ee

\be
\label{6.4}
\frac{dv}{dp}= -\frac{(2K^2+v(p)+4K^2v(p))\left( m_0^2+u(p)\left( 6K^2-12K^2v(p)-v^2(p)\right) \right) }{4K^2(1+6K^2)u(p)}.
\ee

The exact solution is written

\be
\label{6.5}
v(p)=\frac{-2K^2}{1+4K^2},
\ee

\be
\label{6.6}
u(p)= 1-\overline{\Omega}_0 + \overline{\Omega}_0 \exp {-\left[ (3+\frac{4K^2}{1+4K^2})p\right] }.
\ee
Let us note that $ p=0 $ corresponds to $ z=0 $.

There are two hyperbolic fixed points

\be
\label{6.7}
\left( u_1= \frac{4m_0^2}{3(3+16K^2)}\quad , \quad v_1=\frac{3}{2}\right), 
\ee

\be
\label{6.8}
\left( u_2=\frac{(1+4K^2)^2 m_0^2}{2K^2(3+34K^2+96K^2)}\quad , \quad v_2=\frac{-2K^2}{1+4K^2}\right).
\ee
The eigenvalues of the associated linear dynamical system allow the classification of these fixed points.

The first one, $(u_1,v_1)$, is a saddle point (two eigenvalues of opposite sign) which does not concern our solution.The second, $(u_2,v_2)$, is a stable proper node (two identical negative eigenvalues), and the orbit of the exact solution (\ref{6.6}, \ref{6.5}) is   
the horizontal line of the phase portrait which approaches this node.

\begin{figure}[p]
\includegraphics[scale=1]{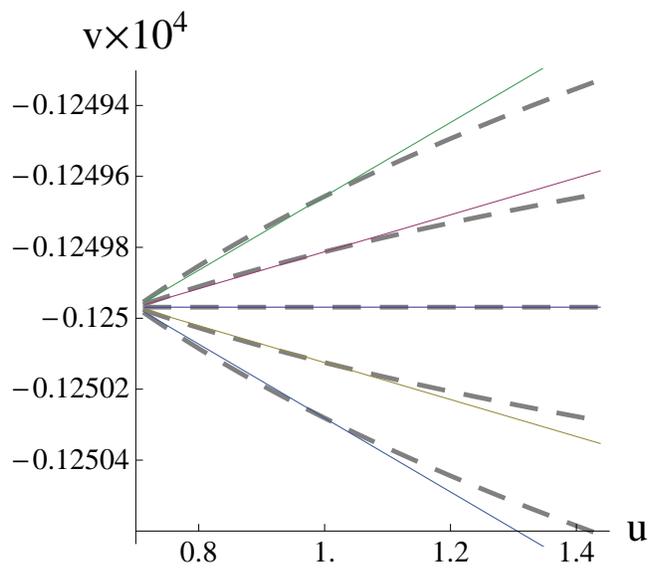}
\caption[]{ Local phase portrait around the horizontal orbit of the exact solution (dashed lines). The continuous lines are the phase portrait of the linearised system.  $ K=1/400 $, $ \Omega_{m,0}=0.3 $.}
\label{fig6}
\end{figure}

The stable node ensures the stability of the exact solution. Let us illustrate this result.
By taking initial conditions on $ v(0) $ in the neighborhood of $v_2= \frac{-2K^2}{1+4K^2} $, (always $ u(0)=1 $ by definition), we have given numerically the local phase portrait around the horizontal orbit of the solution (\ref{6.6}, \ref{6.5}) (cf figure \ref{fig6}) .

\section{Conclusions}

We have obtained  an exact and simple solution $ H(z), \Phi(z) $ for an attractive scalar-tensor model, considered several times previously in the literature. If the constraints of the Solar System observations are respected ($ K^2 \leq (1/400)^2 $), $ H(z)$ cannot be distinguished from $ H_{\Lambda}(z)$ up to a very  large $ z $ for the 
same $ \Omega_{m,0} $. 
But the equation of state  $ w_{DE}(z) $ is different from the scaling solution $ w_{DE}=-1 $ of the $\Lambda CDM$ model since it presents a very slight crossing of the phantom divide line $ w=-1 $ in the neighborhood of $ z=0 $ and tends toward $(4/3)K^2  $ at infinity.

The dependence of the solution $ H/H_{0}$ in Eq.~(\ref{4.19})  on the two parameters $ \Omega_{m,0} $ and $ K^2 $ could be fitted with the supernovae experimental data  using the maximum likelihood method, which involves the minimization of the function
$$
 \chi^2(\Omega_{m,0}, K^2) = 
 \sum_{i=1}^N\left(\frac{D_L(z_i)_\mathrm{obs} - D_L(z_i)}{ \sigma_i}\right)^2
$$
where $ N $ is the number of the observed SNe luminosity distances $ D_L(z_i)_\mathrm{obs} $, $ \sigma_i $ the corresponding errors and $ D_L(z)= (1+z)/H_0\int_{0}^{z}dz'/h(z') $ is the theoretical luminosity distance.
 This fit would be at least as good as the fit of the   $\Lambda$CDM 
model which minimizes $ \chi^2(\Omega_{m,0},0)$ since  the  $\Lambda$CDM solution is obtained for $ K^2\rightarrow 0 $. It would be interesting to verify if this fit respects the solar condition  $K^{2}\leq(1/400)^2$.

Finally, since the  history of the expansion of this model is virtually identical to that of  $\Lambda$CDM, let us see if the observation of the growth of the perturbations could distinguish between these two models. The equation of the fractional density perturbation $ \delta_m(z) $ of this model \cite{beps} and of its  $\Lambda$CDM counterpart can be written

\be
\label{7.1}
u(z) \delta_m''(z)+ \left( \frac{u'(z)}{2}-\frac{u(z)}{1+z}\\\right)   \delta_m'(z) \simeq \frac{3}{2}(1+z)\frac{\Phi^2_0(z)}{\Phi^2(z)}\left( \frac{1+8K^2}{1+6K^2}\right) \Omega_{m,0}\delta_m(z), 
\ee

\be
\label{7.2}
u_{\Lambda}(z) \delta_m''(z)+\left( \frac{u_{\Lambda}'(z)}{2}-\frac{u_{\Lambda}(z)}{1+z}\right)  \delta_m'(z) \simeq \frac{3}{2}(1+z)\Omega_{m,0}\delta_m(z). 
\ee
The coefficients of these linear differential equations are known and numerically close since $ u(z) \simeq u_\Lambda(z) $ and         $ \frac{\Phi^2_0(z)}{\Phi^2(z)} =(1+z)^{\frac{4K^2}{1+4K^2}} \simeq 1 $ until a large redshift. So it seems that it would be difficult to discriminate between these two models on the basis of these perturbation equations.

{\bf Acknowledgments}
We would like acknowledge Bernard Linet, Claude Bervillier, Christos Charmousis, Stam Nicolis and David Polarski for numerous discussions and help all along this work.
\newpage

  

\end{document}